\documentclass[twocolumn,a4paper]{article}

\usepackage[utf8]{inputenc}
\usepackage{amssymb}
\usepackage{caption}
\usepackage{graphicx}
\usepackage{subcaption}
\usepackage{subfig}
\usepackage{amsmath}
\usepackage{natbib}
\usepackage{apalike}

\begin{document}

\markboth{X. Cerdá-Company, C.A. Parraga and X. Otazu}{Which Tone-Mapping-Operator Is the Best? A Comparative Study of Perceptual Quality}

\title{Which Tone-Mapping Operator Is the Best? A Comparative Study of Perceptual Quality}
\author{XIM CERDÁ-COMPANY, C. ALEJANDRO PÁRRAGA and XAVIER OTAZU\\
Computer Vision Center,\\ Computer Science Department,\\ Universitat Autónoma de Barcelona,\\ Spain}
\date{\today}

\maketitle

\providecommand{\keywords}[1]{\textbf{\textit{Index terms---}} #1}

\keywords{Tone-Mapping Operators, Psychophysics, High Dynamic Range,  Visual Perception, Human Visual System, Vision,  Image Processing, ANOVA, Euclidean Distance,Pairwise Comparison}

\begin{abstract}
Tone-mapping operators (TMO) are designed to generate perceptually similar low-dynamic range images from high-dynamic range ones. We studied the performance of fifteen TMOs in two psychophysical experiments where observers compared the digitally generated tone-mapped images to their corresponding physical scenes. All experiments were performed in a controlled environment and the setups were designed to emphasise different image properties: in the first experiment we evaluated the local relationships among intensity-levels, and in the second one we evaluated global visual appearance among physical scenes and tone-mapped images, which were presented side by side. We ranked the TMOs according to how well they reproduce the results obtained in the physical scene. Our results show that ranking position clearly depends on the adopted evaluation criteria, which implies that, in general, these tone-mapping algorithms consider either local or global image attributes but rarely both. We conclude that a more thorough and standardised evaluation criteria are needed to study all the characteristics of TMOs, as there is ample room for improvement in future developments.
\end{abstract}

\section{Introduction}

In almost all naturalistic viewing situations, we are immersed in scenes that could be described as High Dynamic Range (HDR), in other words, the energetic difference between the brightest and the darkest patch is much higher than the difference an imaging device can faithfully capture. For instance, the energy ratio between sunlight and starlight is approximately about 100,000,000:1 \cite{FerwerdaDisplay}. If the Human Visual System (HVS) was to linearly represent these extreme differences, it would require a much larger sensitivity range for its retinal sensors (rods and cones) and neural pathways than is achievable within the limitations of biological chemistry. Instead, millions of years of evolution have solved this problem by adapting the sensorial and neural machinery, allowing it to non-linearly convert the large natural intensity range into a much smaller range of about 10,000:1 \cite{ReinhardBook}. This adaptation is possible because of the combined activity of the pupil, rods and cones, horizontal cells, bipolar cells, amacrine cells, ganglionar cells, interplexiform cells, the primary visual cortex, etc. \cite{Snowden}.

\begin{figure*}[t]
    \centerline
    	{\includegraphics[width=15cm]{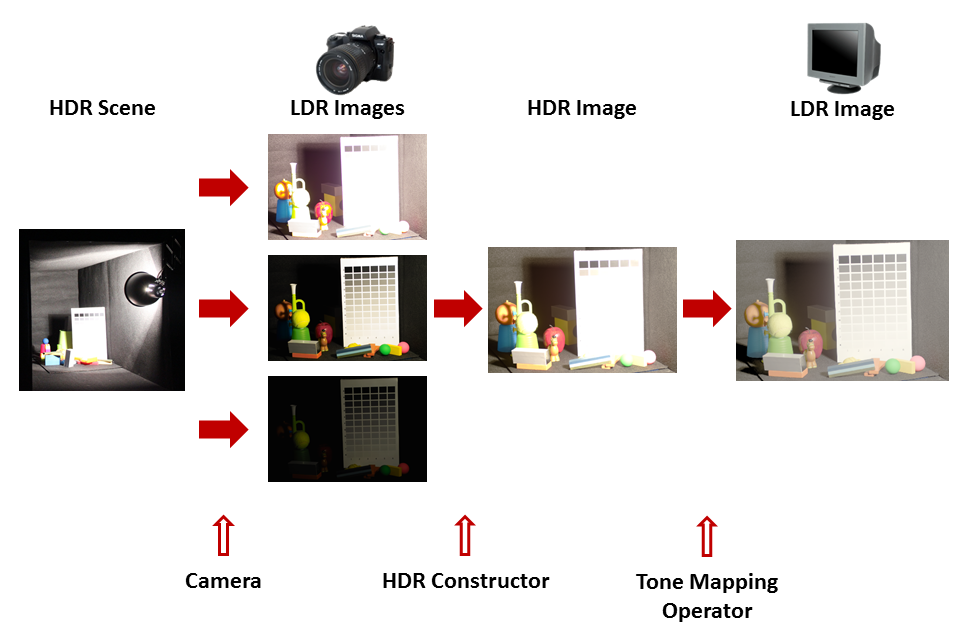}}
    \caption{General workflow of a tone-mapping process. Several low dynamic range (LDR) images, with different exposure times covering different luminosity ranges of a physical scene, are captured by a camera. From these images a high dynamic range (HDR) image is obtained, which can be processed by a TMO in order to obtain a new LDR image that can be displayed on a standard LDR monitor.}
    \label{fig:TMPipeline}
\end{figure*}

This dynamic range presents an important problem for visual representation technologies mostly because common imaging devices (cameras and monitors) are only able to obtain/display images within a much smaller range of about 100:1 \cite{ReinhardBook} which can be increased up to 1000:1 for specialized HDR led-based displays \cite{Ruppertsberg}. To solve this problem, an assortment of non-linear image processing techniques were defined to display HDR scenes in limited Low Dynamic Range (LDR) devices. To construct the HDR image, we use many LDR images of the same scene taken at different exposure values to capture a much larger dynamic range (see Figure~\ref{fig:TMPipeline}). This HDR image is generated by extracting from each LDR image the information corresponding to its region of interest (where it is neither over- or under-exposed) and combining them. Since this new HDR image cannot be displayed on a standard LDR monitor, an algorithmic solution is needed to reduce its dynamic range to match that of the monitor. A common solution is to use a Tone-Mapping Operator (TMO) to reduce the dynamic range while keeping approximately constant some of the original image characteristics. The performance of these TMOs depends on several factors including lighting and viewing conditions, aesthetic/realistic preferences, local/global assumptions, etc. and are usually evaluated using computational (\cite{Aydin,objectiveQuality}) and psychophysical (\cite{DragoExperiment,Kuang2004,Yoshida2005,Ledda,AshikhminExperiment,Cadik2006,Yoshida2007,Kuang2007,iCAM06,Akyuz,Cadik2008}) methods. In this work we present a new set of experiments and analysis to psychophysically evaluate the performance of 15 state-of-the-art TMOs. Unlike previous studies, all the experiments were performed in a controlled environment and tone-mapped images were presented side by side with the physical scene. All tone-mapped images' results were compared to physical scene's results, so we obtained a set of rankings, according to the different criteria, that determine which is the best TMO at representing the original scene as human observers perceive it.

\section{State-of-the-Art}
\subsection{Previous TMO Psychophysical Studies}
Although the idea of using algorithms to match the brightness of real scenes to that of imaging devices was not new \cite{Miller1984,TumblinAndRushmeier}, TMOs did not become popular until the early 2000s due to digital cameras were very exclusive, so many TMOs were developed between 1996 and 2008 \cite{FerwerdaTMO,Ashikhmin,Durand,Fattal,ReinhardTMO,DragoTMO,Krawczyk,Li,ReinhardDevlin,iCAM06,Mertens,Meylan,KimKautz,Ferradans,Otazu}. Thus, up to now many different psychophysical studies have been performed and they can be classified as follows:

\subsubsection{Experiments without HDR scene as reference}
One of the first psychophysical experiments to evaluate TMOs was performed by Drago et al. \citeyear{DragoExperiment}. They compared the performance of 6 TMOs on 4 different (synthetic and photographic) scenes by asking subjects to make pairwise perceptual evaluations and by rating stimuli with respect to three attributes: apparent image contrast, apparent level of detail, and apparent naturalness. Their results showed that preferred operators produced detailed images with moderate contrast.

Kuang et al. \citeyear{Kuang2004} performed pairwise comparisons on 8 different tone-mapping operators using 10 different scenes and two conditions (color and grey-level) where subjects had to choose the preferred image considering general rendering performance (including tone compression performance, color saturation, natural appearance, image contrast and image sharpness). Their results showed that the grey-scale tone-mapping performances are consistent with those in the overall rendering results, if not the same.

\subsubsection{Experiments with HDR scene as reference}
Yoshida et al. \citeyear{Yoshida2005} conducted a psychophysical experiment based on a direct comparison between the appearance of real-world scenes and TMO images of these scenes displayed on a LDR monitor. In their experiment, they differentiate between global and local operators, and introduced, for the first time, the comparison between tone-mapped image and real scene, selecting two different indoor architectural scenes. Fourteen subjects were asked to give ratings according to several criteria like realism (image naturalness in terms of reproducing the overall appearance of the real world views) and image appearance (brightness, contrast, detail reproduction in dark regions, and in bright ones). They found that none of these image appearance attributes had a strong influence on the perception of naturalness by itself.

The same authors extended their previous work with the aim of finding out which attributes of image appearance accounted for the differences between tone-mapped images and the real scene~\cite{Yoshida2007}. They observed a clear distinction between global and local operators. However, they concluded again that none of the image attributes evaluated had a strong influence on the perception of naturalness by itself which suggested that naturalness depends on a combination of the other attributes with different weights.

In an another work, Ashikhmin and Goyal~\citeyear{AshikhminExperiment} performed three different experiments. Subjects ranked different tone-mapped images depending on the asked question. In the first one, the authors asked which image they liked more without having the reference scene. In the second one, the authors asked which image seemed more real without viewing the reference scene and, in the third one, they asked which image was the closest to the real scene viewing the reference scene. They observed that rankings were totally different when subjects could compare the tone-mapped image to the reference scene.

In a subsequent study, Kuang et al. \citeyear{Kuang2007} performed three different experiments they named \textit{preference evaluation}, \textit{image-preference modelling} and \textit{accuracy evaluation}. In the \textit{preference evaluation} experiment, pairwise comparisons between tone-mapped images were performed. Here they used only color images and the aim was to evaluate the general rendering performance by instructing observers to consider perceptual attributes such as overall impression on image contrast, colorfulness, image sharpness, and natural appearance. In contrast, in the \textit{image-preference modelling} experiment, they rated grey-scale images (which were grey-scale versions of the first experiment color images). Here, observers considered perceptual attributes such as highlight details, shadow details, overall contrast, sharpness, colorfulness and appearance of artifacts, comparing the TMO's visual rendering "to their internal representation of a ‘perfect’ image in their minds" \cite{Kuang2007}. In the \textit{accuracy evaluation}, both pairwise comparison and rating techniques were used in order to evaluate the perceptual accuracy of the rendering algorithms. The pairwise comparison of TMOs was performed without viewing the real scene and subjects were asked to compare the overall impression on image contrast, colorfulness, image sharpness, and overall natural appearance. An additional rating evaluation was performed using the real scenes set up in the adjoining room as references. Here, subjects had to rate image attributes like highlight contrast, shadow contrast, highlight colorfulness, shadow colorfulness, overall contrast and the overall rendering accuracy comparing to the overall appearance of the real-world scenes. In both experiments, observers did not have immediate access to the real scene and had to rely on their memories (either short- or long-term) to perform the tasks.

In 2007, Kuang et al. presented the iCAM06 operator \cite{iCAM06}. To validate it, the authors performed two psychophysical experiments similar to the previous ones in \cite{Kuang2007}. The first experiment was a pairwise comparison without viewing the reference scene. Observers had to choose the tone-mapped image that they preferred based on overall impression on image quality (considering contrast, colorfulness, image sharpness, and overall natural appearance). In the second experiment, observers were also asked to evaluate overall rendering accuracy by comparing the overall appearance of the rendered images to their corresponding real-world scenes, which were set up in an adjoining room.

While looking for a definition of an overall image quality measure, Cadík et al. \citeyear{Cadik2006} studied the relationships between some image attributes such as brightness, contrast, reproduction of colors and reproduction of details. They performed two psychophysical experiments, using 14 TMO, in order to propose a scheme of relationships between these attributes, being aware that some special attributes, which were not evaluated (e.g. glare stimulation, visual acuity and artifacts), can influence their relationships. In the first one, 10 subjects were asked to perform ratings using five criteria: overall image quality and the four basic attributes (brightness, contrast, reproduction of detail and colors). These evaluations were performed using a real scene as a reference (a typical real indoor HDR scene). In the second experiment, subjects did not have access to the real scene and they had to rank image printouts according to the overall image quality and the four basic attributes.

In a new study, Cadík et al. \citeyear{Cadik2008} performed exactly the same type of experiments adding two new scenes, that is, they had a total of three scenes, i.e. a real indoor HDR scene, a HDR outdoor scene and a night urban HDR scene. In the first experiment, subjects were asked to rate overall image quality and the quality of reproduction of five attributes by comparing samples to the real scene. These attributes were the same four basic ones of their previous work and the lack of disturbing image artifacts (which was one of the non-evaluated special attributes in \cite{Cadik2006}). These experiments were set-up in an uncontrolled natural environment, so subjects had to perform the experiments at the same time of the day as the HDR image was acquired. In the second experiment, subjects had no possibility of directly comparing to the real scene and had to rank the image printouts according to the overall image quality, and the quality of basic attributes.

\subsubsection{Experiments using an HDR monitor}
In 2005, Ledda et al. \citeyear{Ledda} performed two different psychophysical experiments comparing 6 different tone-mapping operators to linearly mapped HDR scenes displayed on a HDR device. They used 23 different color and grey-scale HDR scenes showing 3 different images per comparison: the HDR and two tone-mapped images. In the first experiment, subjects were asked to select the TMO image more similar to the HDR reference by judging its global appearance. In the second one, they were asked to make their judgment based on reproduction detail.

In a later work, Akyüz et al. \citeyear{Akyuz} asked subjects to rank six images (1 HDR image, 3 tone-mapped images, 1 objectively good LDR exposure value and 1 subjectively good LDR exposure value) according to their subjective preferences. Surprisingly, they found that participants did not systematically preferred tone-mapped HDR images over the best single LDR exposures.\\

As seen above, all previous studies have been focused on subjective comparisons of global and local image appearance attributes such as contrast, colourfulness, sharpness, reproduction artifacts, etc. either within TMOs or against the real scene. While this is no doubt extremely important, we believe a good TMO is also likely to keep the interrelations between objects attributes inside a scene the same as in the physical scene. As far as we know, no study has been conducted to evaluate whether objects represented within a TMO image maintain the same perceived visual differences as the real scene,  and this is the principal aim of our work.

\subsection{Tone-Mapping Operators}
Tone-mapping operators can be classified in two groups: global and local processing. Global operators perform the same computation in all pixels, regardless of spatial position, which make them more computationally efficient at the cost of losing contrast and image detail. Some examples of global TMO are \cite{DragoTMO,FerwerdaTMO,KimKautz,ReinhardDevlin}. On the other hand, local operators, which take into account neighbouring pixels, produce images with more contrast and higher level of detail, but they may show problems with halos around high contrast edges. Local operators are inspired on the local adaptation process present at the early processing stages of the human visual system. Some examples of local operators are \cite{Ashikhmin,Durand,Fattal,Krawczyk,iCAM06,Li,Meylan,Otazu}. There is a third group of “hybrid” tone-mapping operators which could be global or local depending on how they are run. One example is \cite{ReinhardTMO} and another one is \cite{Ferradans}, which is developed in two stages, the first global and the second local. A brief summary of the properties of each tone-mapping operator used in our experiments is given in Table~\ref{table:TMOs} (and in the text below). The first column shows the names that we will use to refer to each operator throughout this work as defined below.

\begin{table}[ht]
    \centering
    \resizebox{\columnwidth}{!}{%
    \begin{tabular}{ccccc}
        \hline
        TMO & Global/Local & HVS & Luminance & Color \\
        \hline
        \textit{Ashikhmin}       & L & \checkmark & \checkmark  &             \\
        \textit{Drago}           & G &           & \checkmark  &             \\
        \textit{Durand}          & L &           & \checkmark  & \checkmark   \\
        \textit{Fattal}          & L &           & \checkmark  &             \\
        \textit{Ferradans}       & H & \checkmark & \checkmark  & \checkmark   \\
        \textit{Ferwerda}        & G & \checkmark & \checkmark  & \checkmark   \\
        \textit{iCAM06}          & L & \checkmark & \checkmark  & \checkmark   \\
        \textit{KimKautz}        & G &           & \checkmark  &             \\
        \textit{Krawczyk}        & L &           & \checkmark  &             \\
        \textit{Li}              & L &           & \checkmark  &             \\
        \textit{Mertens}         & - &           &            &             \\
        \textit{Meylan}          & L & \checkmark & \checkmark  & \checkmark   \\
        \textit{Otazu}           & L & \checkmark & \checkmark  &             \\
        \textit{Reinhard}        & G &           & \checkmark  &             \\
        \textit{Reinhard-Devlin} & G &           & \checkmark  &             \\
        \hline
    \end{tabular}
    }
    \caption{Summary of used TMO's characteristics. Second column shows whether the TMO is global (G), local (L) or hybrid (H). Third column shows whether it is inspired by the Human Visual System, and following columns show whether it processes luminace and color information.}
    \label{table:TMOs}
\end{table}

- \cite{Ashikhmin} (\textit{Ashikhmin}).
This local tone-mapping operator is inspired by the processing mechanisms  present at the first stages of the Human Visual System. Intensity range is compressed by a local luminance adaptation function and, in a last step, detail information is added.

- \cite{DragoTMO} (\textit{Drago}).
This global tone-mapping operator is based on luminance logarithmic compression that, depending on scene content, uses a predetermined logarithmic basis to preserve contrast and details.

- \cite{Durand} (\textit{Durand}).
This local tone-mapping operator decomposes the image in two layers: the base and the detail. The base layer is compressed using bilateral filtering, and the magnitudes of the detail layer are preserved.

- \cite{Fattal} (\textit{Fattal}).
This local tone-mapping operator manipulates the gradient fields of the luminance image. Its idea is to identify high gradients in different scales and attenuate their magnitudes, while maintaining their directions.

- \cite{Ferradans} (\textit{Ferradans}).
This hybrid tone-mapping operator is divided in two stages. In the first stage, it applies a global method that implements the visual adaptation, trying to mimic human cones’ saturation. In the second stage, it enhances local contrast using a variational model inspired by color vision phenomenology.

- \cite{FerwerdaTMO} (\textit{Ferwerda}).
This global tone-mapping operator is based on computational model of visual adaptation that was adjusted to fit psychophysical results  on threshold visibility, color appearance, visual acuity, and sensitivity over the time.

- \cite{iCAM06} (\textit{iCAM06}).
This local tone-mapping operator is based on the iCAM06 color appearance model, which gives the perceptual attributes of each pixel, like lightness, chromaticity, hue, contrast and sharpness. It includes an inverse model which considers viewing conditions to generate the result.

- \cite{KimKautz} (\textit{KimKautz}).
This global tone-mapping operator is based on the assumption that human vision sensitivity is adapted to the average log luminance of the scene and that it follows a Gaussian distribution.

- \cite{Krawczyk} (\textit{Krawczyk}).
This local tone-mapping operator is inspired on the anchoring theory \cite{Gilchrist}. It decomposes the image into patches of consistent luminance (frameworks) and calculates, locally, the lightness values.

- \cite{Li} (\textit{Li}).
This local tone-mapping operator is based on multiscale image decomposition that uses a symmetrical analysis-synthesis filter bank to reconstruct the signal, and applies local gain control to the subbands to reduce the dynamic range. 

- \cite{Mertens} (\textit{Mertens}).
This technique fuses original LDR images of different exposure values (exposure fusion) to obtain the final “tone-mapped” image, which avoids the generation of an HDR image. Guided by simple quality measures like saturation and contrast, it selects “good” pixels of the sequence and combines them to create the resulting image.

- \cite{Meylan} (\textit{Meylan}).
This local tone-mapping operator is derived from a model of retinal processing. In a first step, a basic tone-mapping algorithm is applied on the mosaic image captured by the sensors. In a second step, it introduces a variation of the center/surround spatial opponency.

- \cite{Otazu} (\textit{Otazu}).
This local tone-mapping operator is based on a multipurpose human colour perception algorithm. It decomposes the intensity channel in a multiresolution contrast decomposition and applies a non-linear saturation model of visual cortex neurons.

- \cite{ReinhardTMO} (\textit{Reinhard}).
Hybrid global-local tone-mapping operator, that can be executed as global or local. It performs a global scaling of the dynamic range followed by a dodging and burning (local) processes. In our work, this operator was run as global which is its default value in the toolbox.

- \cite{ReinhardDevlin} (\textit{Reinhard-Devlin}).
This global tone-mapping operator uses a model of photoreceptors adaptation which can be automatically adjusted to the general light level.

\section{Methods}
\subsection{Laboratory Setup}
\label{section:laboratorySetup}
Experiments were performed in a dark room, i.e. in a controlled environment, using a ViSaGe MKI Stimulus Generator, and a calibrated Mitsubishi Diamond-Pro\textregistered 2045u CRT monitor side-by-side with the handmade real HDR scene. Both monitor and real scene were setup so that objects in both scenes subtended approximately the same angle and looked similarly positioned to the observer.

\begin{figure}[t!]
    \centering
    \includegraphics[width=5cm]{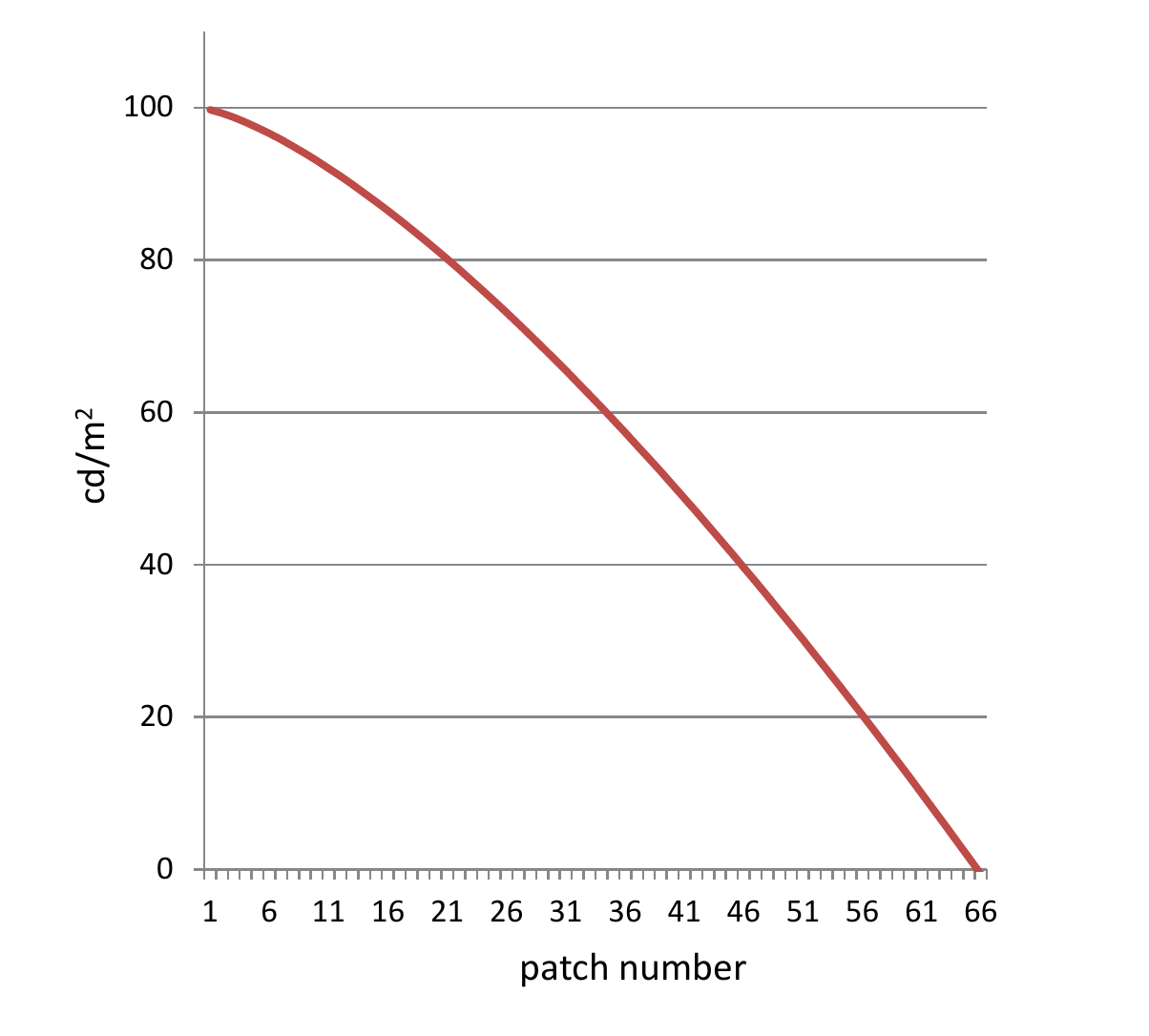}
    \caption{Luminance in $cd/m^{2}$ of Reference Table Patches measured by our PR-655 SpectraScan\textregistered Spectroradiometer. We can observe that their spectral reflectance descreased monotonically from the 1st patch to the 65th.}
    \label{fig:luminance}
\end{figure}

\begin{figure*}[t!]
    \begin{subfigure}[b]{.30\linewidth}
      \centering
      \centerline{\includegraphics[width=5.5cm]{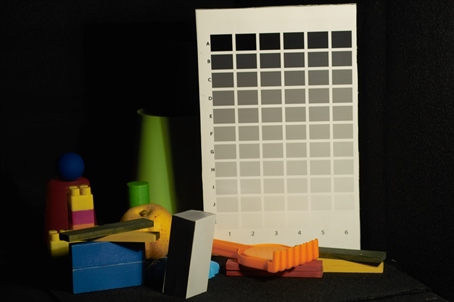}}
      \caption{Scene 1}
    \end{subfigure}
    \hfill
    \begin{subfigure}[b]{0.30\linewidth}
      \centering
      \centerline{\includegraphics[width=5.5cm]{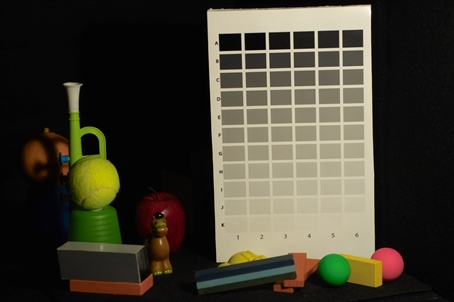}}
      \caption{Scene 2}
    \end{subfigure}
    \hfill
    \begin{subfigure}[b]{0.30\linewidth}
      \centering
      \centerline{\includegraphics[width=5.5cm]{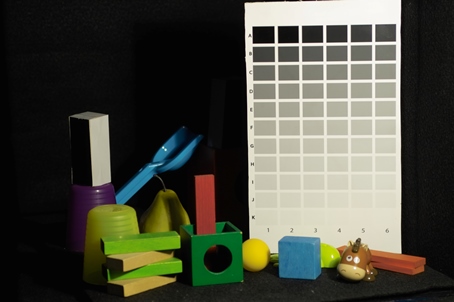}}
      \caption{Scene 3}
    \end{subfigure}
    \caption{To show the general appearance of the physical scenes, here we show a single LDR exposure (choosen by simple visual inspection by the authors) from the set of LDR exposures used to create the HDR images.}
    \label{fig:experimentScenes}
\end{figure*}

We built three different HDR scenes, each including a grey-level reference table and two solid objects (parallelepipeds). The reference table was built by printing a series of 65 grey squares (2.8 x 2.2 cm) arranged in a flat 11x6 distribution whose spectral reflectance increased monotonically from top to bottom, as measured by our PR-655 SpectraScan\textregistered Spectroradiometer (see Figure~\ref{fig:luminance}). The parallelepipeds consisted of 3.6 x 3.6 x variable length between 9.4 and 10 cm pieces of wood, whose sides were covered with random samples of the same printed paper as the reference tables. The rest of the scenes consisted of many plastic and wooden objects of different colours and shapes (see Figure~\ref{fig:experimentScenes}).

We placed one of the parallelepipeds in the bright part of the scene and the other under the shade cast by the reference table. This was done in order to have some reference objects directly lit by the scene illumination and others in the dark. Two surfaces of one parallelepiped and three of the other one were visible from subjects' location, resulting in 15 different grey visible surfaces in total. Scenes showed a wide range of colors and were illuminated with an incandescent lamp of 100W whose bulb was painted blue to simulate D65 illumination.

We photographed the real scene using a calibrated Sigma Foveon SD10 camera placed in the exact same position as subjects’ head during the psychophysical experiment, and camera setup was arranged so that the images presented later on the monitor looked geometrically similar to the real ones shown beside them. The experiments were conducted in a controlled environment, where the only light sources were the lamp illuminating the real scene and the indirect light produced by the CRT monitor. Since experiments were performed in a dark room, reflections from all other objects and the walls were minimised. The dynamic range of the scenes were approximately $10^5$ for scene 1 and $10^6$ for scenes 2 and 3.

We know that an accurate representation of the scene luminance distribution is not possible to achieve from camera images, but multiple exposure values improve the image information contained in the HDR image \cite{McCannRizzi}. Thus, a set of 25 photographs were taken at different exposure values (from 15 sec to 1/6000 sec) using the same aperture, focal distance, zoom settings and visual field. Individual images were stored in RAW format and transformed into 16 bits sRGB (using the camera manufacturer’s software). HDR images were obtained using the HDR Toolbox for MATLAB \cite{Banterle}.

\subsection{Experiments}
In order to compare TMO algorithms, we performed two different experiments. The aim of the first experiment was to study the internal (local) relationships among grey-levels in the tone-mapping image and in the real scene. The aim of the second experiment was to rank TMOs according to how similar their results were perceived to the real scene, using a global criterion. In both cases, we obtained a ranking of the different TMOs.

All experiments started with a 1-minute subject adaptation to the ambient light and implementations of some of TMOs have been obtained from HDR Toolbox for MATLAB \cite{Banterle} (all except~\textit{Ferradans},~\textit{iCAM06},~\textit{Li},~\textit{Meylan} and~\textit{Otazu}, which have been obtained from their authors' web page). In order to avoid to benefit any of these operators, we have run all of them with their default parameters. In \textit{Ferradans}' case, we had to determine two different parameters and we chose the default values specified in their paper ($\rho=0$ and $\alpha^{-1}=3$).

\subsubsection{Experiment 1: Local Criterion}
\label{Experim:LocalCriterion}

\begin{figure*}[t!]
    \begin{subfigure}[b]{.48\linewidth}
      \centering
      \centerline{\includegraphics[width=4.0cm]{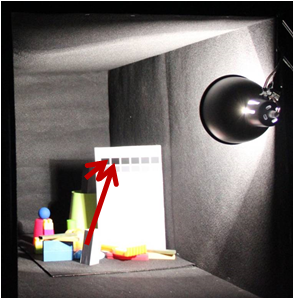}}
      \caption{Task 1}
      \label{Experim1:step1}
    \end{subfigure}
    \hfill
    \begin{subfigure}[b]{0.48\linewidth}
      \centering
      \centerline{\includegraphics[width=6.0cm]{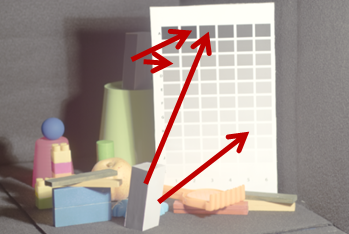}}
      \caption{Task 2}
      \label{Experim1:step2}
    \end{subfigure}
    \caption{In Experiment 1, observers performed two tasks. In \textit{Task 1} (Figure~\ref{Experim1:step1}), observers had to match the brightnesses of the 5 parallelepipeds' surfaces to the brightnesses of 5 patches in the reference table. In \textit{Task 2} (Figure~\ref{Experim1:step2}), observers had to perform the same task on the TMO image displayed on the calibrated monitor. (Red arrows are randomly drawed just for illustrative purposes).}
    \label{Experim1:pipeline}
\end{figure*}

This experiment consists on two different tasks:

\textit{Task 1}.
After adaptation, subjects were asked to match, in the real scenes (i.e. with monitor turned off), the brightnesses of the 5 parallelepipeds' surfaces to the brightnesses of the reference table in the scene (see Figure~\ref{Experim1:step1}). Although there were no time constraints to perform the tasks, they were advised to take no more than 30 seconds per match.

\textit{Task 2}.
Here the real scene was not visible and the observers just saw a digital tone-mapped version of the real scene presented on the monitor. Their task was similar as in \textit{Part 1}, except that the matchings were conducted entirely on the grey patches shown on the screen (see Figure~\ref{Experim1:step2}).\\

There were three conditions for the experiment, corresponding to the three different scenes created (see Figure~\ref{fig:experimentScenes}). Observers matched the 5 surfaces in all 15 different tone-mapped images and in the real scene for each of the three scenes (conditions), so they performed a total of 240 matchings. Matchings were conducted by writing results on a piece of paper. For every scene, all tone-mapped images were shown in random order. 

\textit{Task 1} was completed by a group of 12 observers with normal or corrected-to-normal vision, recruited from our lab academic/research community. This group (8 males and 4 females) was comprised by people aged between 17 and 54. Nine of them were completely naive to the aims of the experiment. 
\textit{Task 2} was completed by 10 of the previous observers (8 males and 2 females).

After all matches, we converted the results to $cd/m^{2}$ using the measurements shown in Figure~\ref{fig:luminance}.

\subsubsection{Experiment 2: Global Criterion}
\label{Experim:GlobalCriterion}

This experiment consisted on a pairwise comparison of tone-mapped images obtained using different TMOs in the presence of the original scene (side by side). After 1-minute adaptation in front of the physical scene, a pair of tone-mapped images of the same physical scene was randomly selected and presented sequentially to the observer on the CRT screen (besides the real scene). Objects in both digital and physical scenes looked geometrically the same, subtending the same angle to the observer. Subjects could toggle a gamepad button to select which image of the tone-mapped pair was presented on the monitor (only one image was displayed at a time) and to select the image that was more similar to the real scene. As before, there was no time limit to perform the comparisons, but subjects were advised to complete a trial in less than 30 seconds. After an image was chosen, a grey background was shown for two seconds, and a different random pair was selected for the next trial. Every subject performed 105 comparisons per scene taking around 25 minutes in total. There were three experimental conditions, corresponding to the three different physical scenes created (see Figure~\ref{fig:experimentScenes}). Between conditions, subjects were forced to go out of the laboratory to take a 5 to 10 minutes break while the physical scene was replaced.

A group of 10 people with normal or corrected-to-normal vision, 7 males and 3 females recruited from our lab academic and research community, completed this experiment. This group was comprised by people aged between 17 and 54. Seven of them were naive to the aims of the experiment.

\section{Results}
\subsection{Experiment 1: Local Criterion}
\label{Experim1:results}
In every scene evaluated in Section~\ref{Experim:LocalCriterion}, we obtained 16 sets of experimental values (the real scene in \textit{Task 1} plus 15 tone-mapped images in \textit{Task 2}) for each of the 15 considered surfaces.

We performed two different analysis to evaluate to what extend the local interrelations perceived by the observers in the tone-mapped versions corresponded to those perceived in the real scene.

\subsubsection{Analysis 1}
\label{paragraph:Analysis1}
In the first analysis, we compared the perceived distances from each tone-mapping algorithm to the real scene. The real scene and the TMOs were defined in a 15-dimensional space, where each dimension corresponds to each evaluated surface. To obtain this 15-dimensional vector, for each TMO and for the real scene, the psychophysical data for all observers for each grey-level surface was averaged. Then, in order to avoid the distortion and other noise, we performed a Principal Component Analysis (PCA) to reduce the dimensionality and we kept the first 6 components (which represents 95.40\% of data information). A ranking was obtained (see Table~\ref{table:PCA}) by measuring the Euclidean distance from each operator to the real scene in this 6-dimensional space.

\begin{table}[ht]
        \centering
        \begin{tabular}{ccc}
            \hline
            \multicolumn{3}{c}{Euclidean Distance} \\
            \hline
            TMO                     & Distance  & Type \\
            \hline
            \textit{iCAM06}         & 2.26 & Local \\
            \textit{Durand}         & 4.26 & Local \\
            \textit{Fattal}         & 4.52 & Local \\
            \textit{Li}             & 4.85 & Local \\
            \textit{Mertens}        & 4.89 & - \\
            \textit{KimKautz}       & 4.96 & Global \\
            \textit{Krawczyk}       & 5.16 & Local \\
            \textit{Reinhard}       & 5.24 & Global \\
            \textit{Meylan}         & 5.27 & Local \\
            \textit{Reinhard-Devlin}& 5.29 & Global \\
            \textit{Ferwerda}       & 5.66 & Global \\
            \textit{Ferradans}      & 5.91 & Hybrid \\
            \textit{Drago}          & 6.02 & Global \\
            \textit{Otazu}          & 6.22 & Local \\
            \textit{Ashikhmin}      & 9.00 & Local \\
            \hline
        \end{tabular}
        \caption{A PCA was performed on the grey-level surfaces' psychophysical data in order to reduce its dimensionality. In the space defined by the six first principal components (representing the 95.40\% of data information), we compute the perceptual Euclidean distance from each evaluated TMO to the real scene.}
        \label{table:PCA}
\end{table}

\subsubsection{Analysis 2}
\label{paragraph:Analysis2}
\begin{figure*}[ht]
\centering
\includegraphics[width=\textwidth]{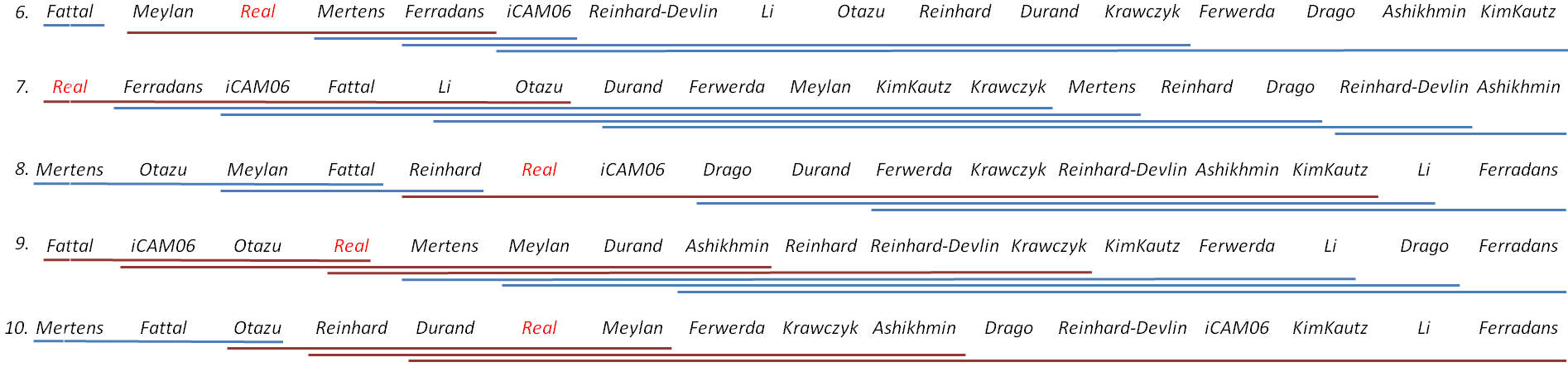}
\caption{Interesting cases of Fisher's Least Significant Difference Post-hoc Test carried out for the different grey-level surfaces (from 6th to 10th). Each line corresponds to the set of TMOs that are not significantly different at $p<0.05$. The real scene is indicated in red.}
\label{fig:scene2FLSD}
\end{figure*}

In order to know if there are significant differences between these sets of experimental values, we calculated 15 different ANOVAs (one per grey-level surface). In all cases, significant differences (at $p<0.05$) between the algorithms were found. For each ANOVA, a Fisher’s Least Significant Difference (FLSD) post-hoc test was computed (Figure~\ref{fig:scene2FLSD}) to analyze which TMO was different to the real scene. In this figure, rows correspond to grey-level surfaces and for each one we show the several groups obtained by FLSD. Every group shows the set of algorithms that could be considered similar to each other. In each row, algorithms' names are arranged according to their output's similarity to the real scene: the further from the “real scene” label (in red), the more dissimilar the results. From this data, a ranking (Table~\ref{table:ANOVA}) of tone-mapping operators is obtained by computing how many times a TMO can be considered similar to the real scene. Data obtained in this section is shown in Appendix~\ref{Appendix:statisticalResults}.

\begin{table}[ht]
        \centering
        \begin{tabular}{ccc}
            \hline
            \multicolumn{3}{c}{ANOVA} \\
            \hline
            TMO                     & Score  & Type \\
            \hline
            \textit{iCAM06}         & 14 & Local \\
            \textit{Ferradans}      & 11 & Hybrid \\
            \textit{KimKautz}       & 10 & Global \\
            \textit{Durand}         & 9 & Local \\
            \textit{Fattal}         & 9 & Local \\
            \textit{Krawczyk}       & 9 & Local \\
            \textit{Li}             & 9 & Local \\
            \textit{Mertens}        & 9 & - \\
            \textit{Reinhard-Devlin}& 9 & Global \\
            \textit{Drago}          & 8 & Global \\
            \textit{Meylan}         & 8 & Local \\
            \textit{Otazu}          & 8 & Local \\
            \textit{Reinhard}       & 8 & Global \\
            \textit{Ashikhmin}      & 7 & Local \\
            \textit{Ferwerda}       & 6 & Global \\
            \hline
        \end{tabular}
        \caption{Number of times (score) that each evaluated TMO can be considered similar to the real scene (over 15).}
        \label{table:ANOVA}
\end{table}

\subsection{Experiment 2: Global Criterion}
\label{sec:resultsExperiment2}
From the pairwise comparison results, we defined a preference matrix for each subject and each scene. We constructed a directed graph where the nodes were the evaluated TMOs and the arrows pointed from a preferred TMO to a non-preferred TMO, e.g. if the TMO$_{i}$ is preferred over the TMO$_{j}$ (tone-mapped image from TMO$_{i}$ is more similar to the real scene than the one from TMO$_{j}$), we drew an arrow from node$_{i}$ to node$_{j}$, for $i \neq j$.

From this graph, we were able to analyse intra-subject consistency coefficient $\zeta$ for each scene. The consistency coefficient for each subject and scene is defined by

\begin{equation}
    \begin{split}
        d_{st} =& \frac{n(n-1)(2n-1)}{12}-\frac{1}{2}\sum_{i=1}^{n}a_{ist}^2,\\
        \zeta_{st} =& 
            \begin{cases}
                1-\frac{24d_{st}}{n^3-n}, \textrm{if $n$ is odd}.\\
                1-\frac{24d_{st}}{n^3-4n}, \textrm{if $n$ is even}.
            \end{cases}
    \end{split}
\label{eq:consistencyCoeff}
\end{equation}

\noindent
where $s$ is the scene number ($s\in[1,3]$), $t$ is the subject number ($t\in[1,m]$), $n$ is the number of evaluated TMOs, and $a_{i}$ is the number of arrows which leave the node$_{i}$. The maximum $\zeta$ value is 1 (perfect consistency within-subject).

The consistency or the agreement between subjects, i.e. inter-subject agreement, is measured by the Kendall Coefficient of Agreement~\cite{Kendall}. This measure is defined by

\begin{equation}
    u_{s} = \frac{2\sum_{i \neq j} \binom{p_{ij}}{2}}{\binom{m}{2}\binom{n}{2}}-1
    \label{eq:agreementMeasure}
\end{equation}

\noindent
where $p_{ij}$ is the number of times TMO$_{i}$ is preferred over TMO$_{j}$ and $m$ is the number of subjects. Since the number of subjects is even ($m=10$), the possible minimum value of $u$, given by Equation~\ref{eq:agreementMeasure}, is $u = -\frac{1}{m-1}$ and its possible maximum value is $u=1$.

In order to study if $u_{s}$ values are significant, we used the chi-sqared test ($\chi^2$). The $\chi_{s}^2$ values are defined by

\begin{equation}
    \chi_{s}^2 = \frac{n(n-1)(1+u_{s}(m-1))}{2}
    \label{eq:chiSquared}
\end{equation}

\noindent
The number of degrees of freedom of the chi-squared test is given by $\frac{n(n-1)}{2}$.

\begin{table}[ht]
    \centering
    \begin{tabular}{c|c|c|cc}
        Scene & $\bar{\zeta}$ & $u$ & $\chi^2$ & $p$, 105 df \\
        \hline
        1 & $0.91$ & $0.61$ & $681$ & $<0.001$\\
        2 & $0.95$ & $0.65$ & $719$ & $<0.001$\\
        3 & $0.93$ & $0.55$ & $624$ & $<0.001$\\
        \hline
    \end{tabular}
    \caption{Summary of all statistical analysis from section~\ref{sec:resultsExperiment2}. Using the preference matrix of each subject and scene, we computed the consistency of intra-subject evaluation (consistency coefficient $\zeta$), i.e. how much consistent the subjects were in their comparisons. Furthermore, we analysed the consistency $u$ of the inter-subjects evaluation (Kendall Agreement Coefficient). We computed the chi-squared test to see if these $u$ values were significant and we saw that they are.}
    \label{table:coefficients}
\end{table}

\begin{figure*}[t!]
    \begin{subfigure}[b]{.329\textwidth}
      \centering
      \centerline{\includegraphics[width=\textwidth]{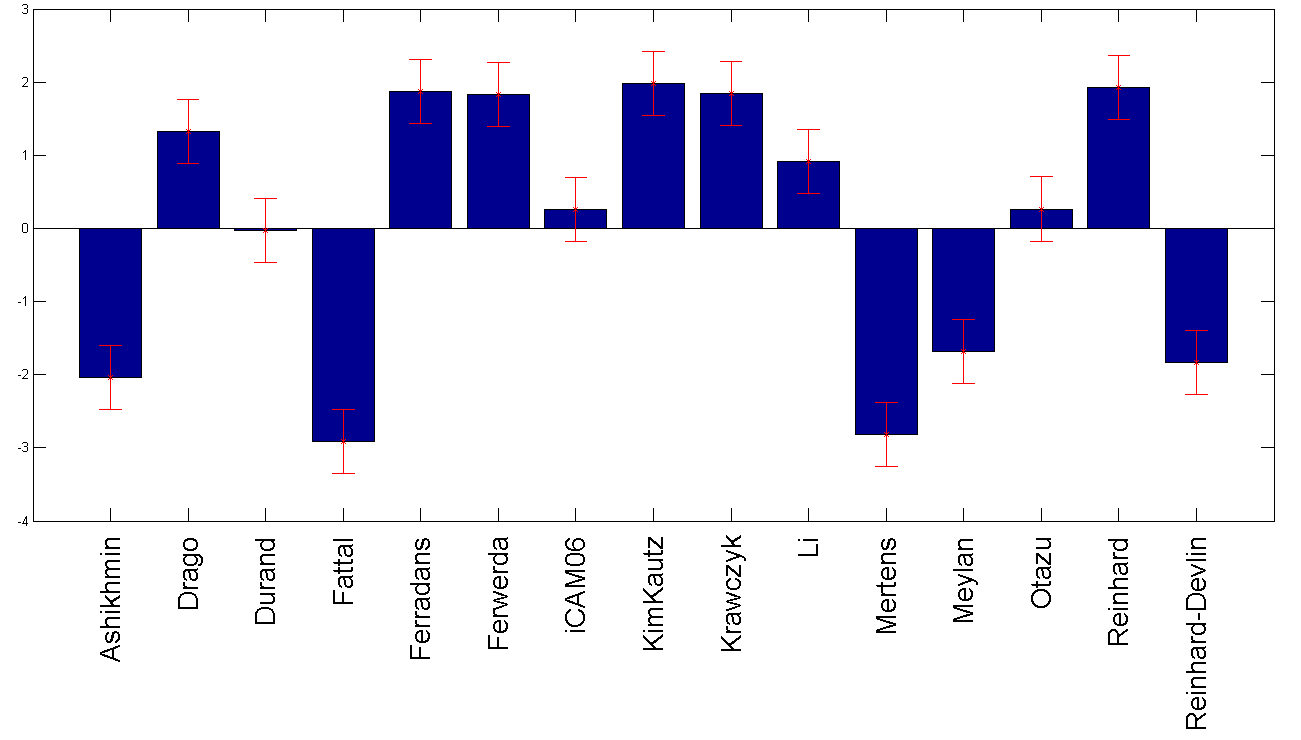}}
      \caption{Scene 1}
      \label{Thurstone:scene1}
    \end{subfigure}
    \hfill
    \begin{subfigure}[b]{0.329\textwidth}
      \centering
      \centerline{\includegraphics[width=\textwidth]{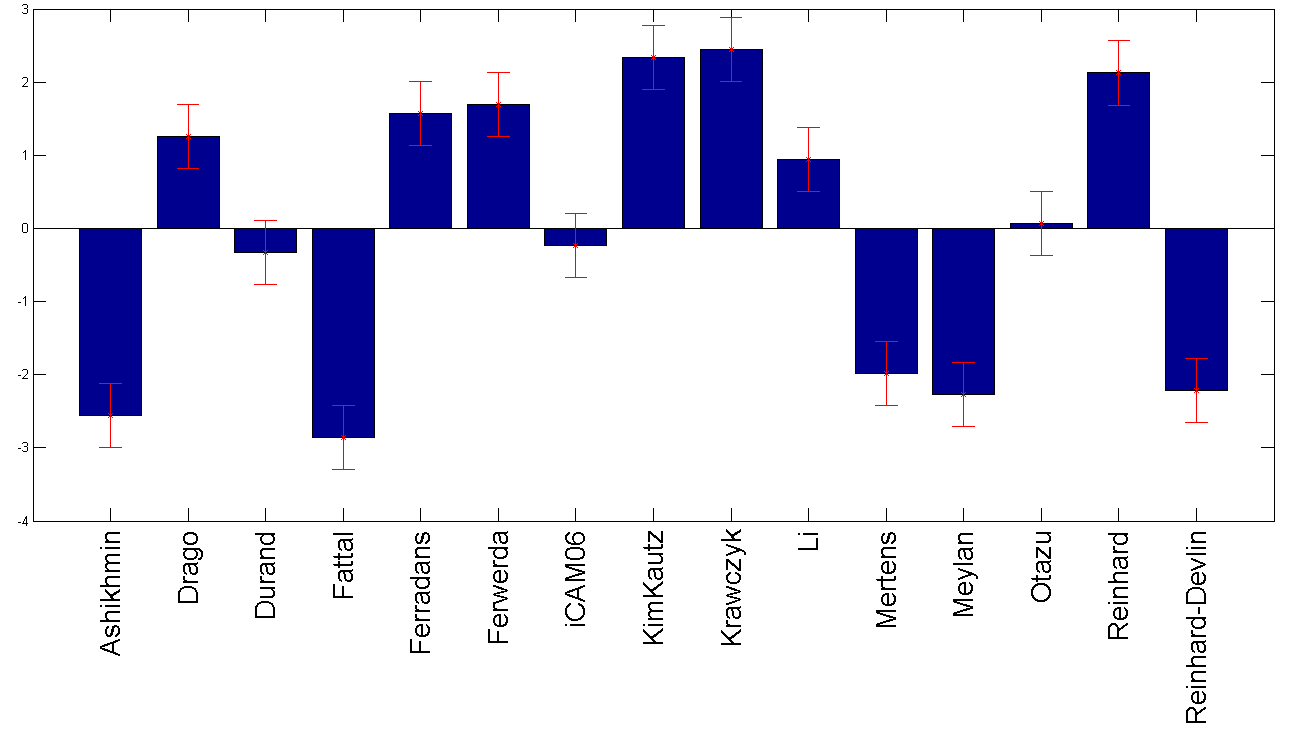}}
      \caption{Scene 2}
      \label{Thurstone:scene2}
    \end{subfigure}
    \begin{subfigure}[b]{0.329\textwidth}
      \centering
      \centerline{\includegraphics[width=\textwidth]{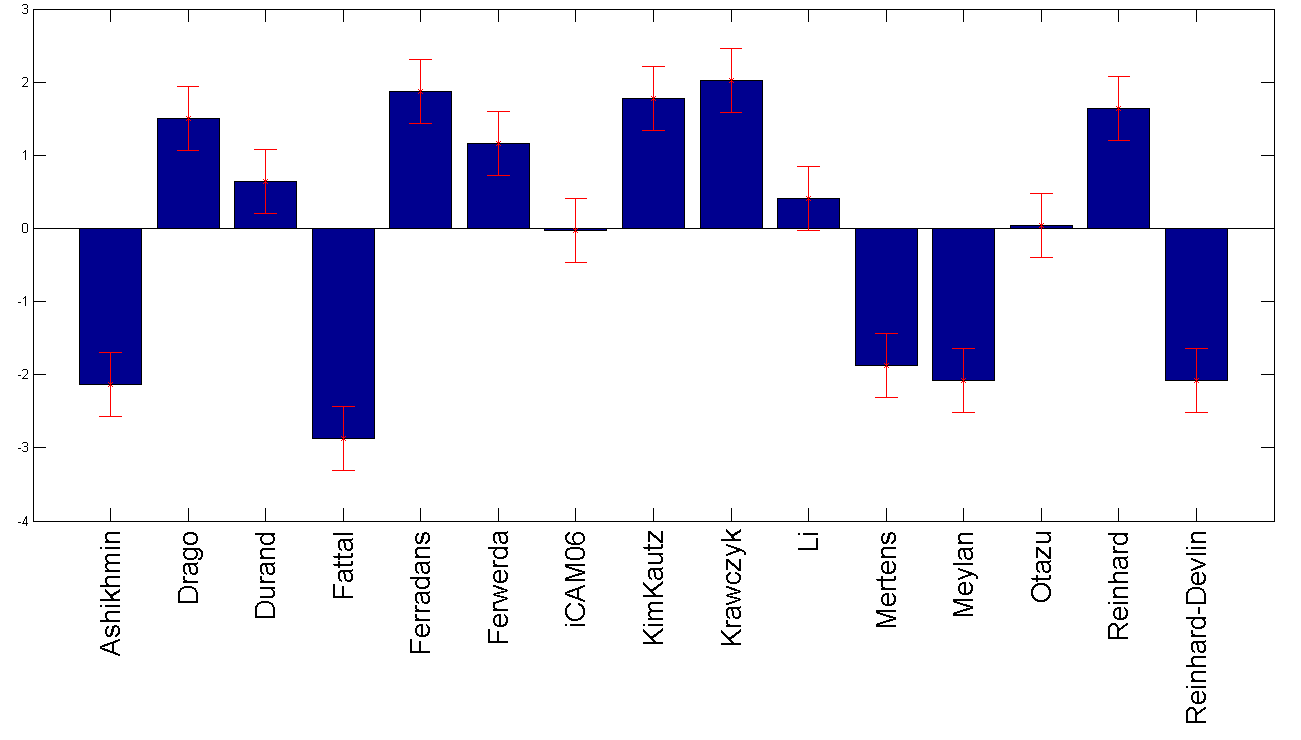}}
      \caption{Scene 3}
      \label{Thurstone:scene3}
    \end{subfigure}
    \caption{Case V Thurstone Law's scores for each evaluated TMO for each different scene.}
    \label{Experim2:Thurstone}
\end{figure*}

In Table~\ref{table:coefficients}, we show all statistical measures for each scene, where we can see that intra- and inter-subject consistency values are very high and they are statistically significant. Then, in Figure~\ref{Experim2:Thurstone}, we show the results of the overall paired comparison evaluations for every scene (obtained from Thurstone's Law of Comparative Judgment, Case V~\cite{Thurstone}) and $95\%$ confidence limits. Since Spearman's correlation between rankings of the three scenes are equal or higher than $0.90$ ($p<0.05$), we computed the mean value along all the scenes (Table~\ref{table:Thurstone}).

\begin{table}[ht]
        \centering
        \begin{tabular}{ccc}
            \hline
            \multicolumn{3}{c}{Averaged Thurstone Law's Scores} \\
            \hline
            TMO                     & Score  & Type \\
            \hline
            \textit{Krawczyk}       & 2.10 & Local \\
            \textit{KimKautz}       & 2.03 & Global \\
            \textit{Reinhard}       & 1.90 & Global \\
            \textit{Ferwerda}       & 1.57 & Global \\
            \textit{Ferradans}      & 1.48 & Hybrid \\
            \textit{Drago}          & 1.36 & Global \\
            \textit{Li}             & 0.75 & Local \\
            \textit{Otazu}          & 0.13 & Local \\
            \textit{Durand}         & 0.10 & Local \\
            \textit{iCAM06}         & 0.00 & Local \\
            \textit{Meylan}         & -2.01 & Local \\
            \textit{Reinhard-Devlin}& -2.04 & Global \\
            \textit{Mertens}        & -2.22 & - \\
            \textit{Ashikhmin}      & -2.25 & Local \\
            \textit{Fattal}         & -2.89 & Local \\
            \hline
        \end{tabular}
        \caption{Ranking obtained by averaging the scores given by Case V Thurstone Law in the three different scenes.}
        \label{table:Thurstone}
\end{table}

\section{Discussion}
From Section~\ref{Experim:LocalCriterion}, we see in Table~\ref{table:PCA} that \textit{iCAM06} has the smallest local distance to the real scene. Similarly, in Table~\ref{table:ANOVA} \textit{iCAM06} is again the best algorithm and can be considered similar to the real scene on 14 out of 15 cases, hence local relationships among grey-levels in \textit{iCAM06} tone-mapped image can be considered similar to the ones in the real scene. Since \textit{iCAM06} is based on a color appearance model that estimates perceptual attributes such as lightness, chromaticity, hue, contrast and sharpness, it is expected for this method to be in line with observers' perception.

Spearman's correlation between rankings of Table~\ref{table:PCA} and Table~\ref{table:ANOVA} shows a correlation of $0.62$ ($p<0.05$), being some algorithms in very different positions, where \textit{Ferradans} is an interesting case. In Table~\ref{table:PCA} this TMO algorithm is in the 12th position, but in Table~\ref{table:ANOVA} it is in the 2nd one. In the first case (Euclidean distance), its distance to the real scene is bigger than most of the other algorithms and in the second case (ANOVA analysis), it can be considered similar to the real scene in 11 out of 15 cases. This is because there are a few instances where \textit{Ferradans} is very different from the real scene and these instances bias the Euclidean distance calculation. This can be seen in Figure~\ref{fig:scene2FLSD}, where \textit{Ferradans}' results are far away from the real scene. An opposite example is \textit{Durand}, which is 6 times dissimilar to the real scene (5 times less than \textit{iCAM06}) in Table~\ref{table:ANOVA} but it is the second best algorithm in Table~\ref{table:PCA}. In contrast to \textit{Ferradans}, when it is not considered similar to the real scene, it is quite close to the real scene's group. Hence, the Euclidean distance is small and it is high ranked.

From Section~\ref{Experim:GlobalCriterion}, it can be seen that some algorithms (\textit{KimKautz}, \textit{Krawczyk} and \textit{Reinhard}) have a good values independently of the scene. We calculated the Spearman’s correlation between the three real scenes and we saw that there is a significant high correlation between the results obtained for the different scenes ($p<0.05$). This correlation is equal or higher than $0.90$. Hence, algorithms have similar behaviour across different scenes.\\

Comparing the results of these two experiments, we observe that in Section~\ref{Experim:LocalCriterion} (Local Criterion experiment - see Tables~\ref{table:PCA} and~\ref{table:ANOVA}), local TMOs are quite better than global ones. Similarly, in Section~\ref{Experim:GlobalCriterion} (Global Criterion experiment - see Table~\ref{table:Thurstone}), global TMOs are quite better than local ones. We have computed the Spearman's correlation between all local and global experiments rankings and we saw that there is no correlation.

An interesting example of different behaviour between local and global experiments is \textit{iCAM06}. In the two local criterion rankings it is in the first position, but in the global rankings it is in a middle position. This means that it correctly reproduces local relationships among grey-levels, but global features are not maintained. An extreme example is \textit{Fattal}, which is in the third and fifth positions in the two local criterion rankings, but is the last in the global criterion ranking. This can be explained if we take into account that \textit{Fattal} is based on local features, e.g. luminance gradients, but it does not enforce global features (such as global brightness and contrast).\\

From the previous results, we infer that global appearance does not only depend on the correct reproduction of local intensity relationships, but it might depend on many other weighted local attributes, such as the reproduction of grey-level and color relationships, contrast, brightness, artifacts, level of detail, etc. This is in agreement with other authors \cite{Yoshida2005,Yoshida2007,Cadik2006,Cadik2008}. Furthermore, our results show that global attributes should also be considered to correctly reproduce global appearance.

Regarding the question of which is the best algorithm, \textit{KimKautz}, \textit{Krawczyk} and \textit{Reinhard} are quite close in Euclidean distance and ANOVA score (Local Criterion) and averaged Thurstone Law's score (Global Criterion), hence all of them can be considered equally good. However, if we look at the relative order position (which is always \textit{KimKautz} and \textit{Krawcyk} in the first or second position and \textit{Reinhard} in the third), we could consider \textit{KimKautz} and \textit{Krawczyk} slightly better than \textit{Reinhard}.

\subsection{Comparison to other Studies}
Since in Section~\ref{Experim:LocalCriterion} we took into account a particular local criterion which, up to our knowledge, has never been studied in this kind of psychophysical experiments, we try to compare our local experiment results to results obtained from other works that study TMOs applied on grey-level images (because our local analysis has been performed on grey-level surfaces).

In contrast, many works perform global comparisons, either with (as in our work) or without the real scene.\\

Although Kuang et al. \citeyear{Kuang2004} performed an experiment without a real scene reference, our global results agree with theirs in that \textit{Fattal} is the worst operator and \textit{Reinhard} is near the best ranked. Contrary to our results, Kuang et al. \citeyear{Kuang2004} say that \textit{Durand} is better than \textit{Reinhard}. The reason could be that they might have used \textit{Reinhard} as a local operator. Furthermore, they performed a study with grey-scales images and their results showed that \textit{Durand} was better than \textit{Reinhard}, but iCAM was worse than \textit{Reinhard}, which is approximately similar to our local experiments results. They differ in iCAM's result, but they used iCAM \cite{iCAM} instead of \textit{iCAM06}, as in our case.

Yoshida et al. \citeyear{Yoshida2005,Yoshida2007} performed experiments with architectural indoor HDR scenes and they concluded that \textit{Reinhard} and \textit{Drago} were good in terms of naturalness and \textit{Durand} was not ranked as highly as in \cite{Kuang2004} (which was without reference). Our results agree with Yoshida et al. \citeyear{Yoshida2005,Yoshida2007}. Moreover, Yoshida et al. \citeyear{Yoshida2007} showed that global and local operators obtain different results, but global TMO results are more similar among themselves than local TMO. This relationship is also present in our study (Tables~\ref{table:PCA},~\ref{table:ANOVA} and~\ref{table:Thurstone}): on one hand we observe that local tone-mapping operators are better than global ones according to the local criterion and on the other hand global operators are much better than local ones considering the global criterion.

Ledda et al. \citeyear{Ledda} used a High Dynamic Range display and obtained a ranking according to the overall similarity of TMO images. In this ranking, iCAM was the first one, which does not agree with our results. In addition, their ranking shows the following TMO’s order: \textit{Reinhard}, \textit{Drago} and \textit{Durand}, which match to our results. Furthermore, these authors also performed experiments in grey-scales obtaining \textit{Reinhard} as the best one, which does not agree with our results.

Cadík et al. \citeyear{Cadik2006,Cadik2008} performed a very exhaustive study of perceptual attributes. We agree with some of their results like the high position of \textit{Reinhard} (nearly the best) and unnaturalness of \textit{Fattal}. Moreover, we strongly agree with them in that the best overall quality is generally observed in images produced by global tone-mapping operators. Nevertheless, we want to point out that there was some conflict between the last two results. In the first one (\cite{Cadik2006}), \textit{Durand} was the worst operator, even worse than \textit{Fattal}, but in the second one (\cite{Cadik2008}), \textit{Fattal} was the worst and \textit{Durand} was in a middle position. Our results goes in line with Cadík et al. \citeyear{Cadik2008}.

We do not agree with Kuang et al. \citeyear{Kuang2007} in that \textit{Durand} is always the best operator, with and without reference. Furthermore, in contrast to our results, \textit{Reinhard} is in a middle position of the ranking. As commented above, \textit{Reinhard} may have been used as a local operator, while we are using it as a global one.

Kuang et al. \citeyear{iCAM06} suggested, again, that \textit{Durand} was better than \textit{Reinhard} and \textit{iCAM06} was even better than \textit{Durand}. In our results, \textit{Durand} and \textit{iCAM06} are quite close, but \textit{Reinhard} is much better than them. Again, \textit{Reinhard} could have been used as a local TMO.

In a similar study as Kuang et al. \citeyear{Kuang2004}, Ashikhmin and Goyal \citeyear{AshikhminExperiment} concluded that, comparing to the real scene, \textit{Fattal} and \textit{Drago} were two of their overall best performers. We do not agree that \textit{Fattal} is one of the best performers, but we have to point out that, in their work, they tuned the TMO's parameters, so this leads to the conclusion that \textit{Fattal} could be a good TMO when a fine tuning of the parameters is performed. Furthermore, in their work, \textit{Drago} obtained more or less the same results as \textit{Fattal}, but \textit{Reinhard} obtained worse results than them. They do not specify how they run \textit{Reinhard}, but it is possible that they run it in the local mode. They obtained that the trilateral filtering \cite{Choudhury}, which is an improvement of \textit{Durand}, was the worst TMO, so it makes sense that, in our work, \textit{Durand} has obtained worse results than \textit{Drago} and \textit{Reinhard}.

In \cite{Akyuz}, the outputs of the most internally sophisticated TMO are statistically worse than the best single LDR exposure. Since a global operator is generally less sophisticated than a local, we could expect that global TMO results are better than local TMO results. Contrary to this theory, \textit{Mertens} (which cannot be considered a sophisticated TMO because it uses single exposure values) is on middle positions in the local experiments but it is one of the worst in the global experiments.\\

To finish this section, literature research leads us to some papers about metrics in order to compare the different tone-mapped images. For example, Ferradans et al. \citeyear{Ferradans} performed an evaluation of several TMOs using the metric of Aydin et al. \citeyear{Aydin}. Although it is not the purpose of our work, we performed a very preliminary analysis comparing our results to the Aydin et al.'s \citeyear{Aydin} results shown in \cite{Ferradans}. We agree that \textit{Fattal} was the operator with highest total error percentages, but we don’t agree with the general overall TMOs ranking. A detailed analysis comparing numerical metrics and psychophysical results is scheduled for future work.

\section{Conclusions}
Our results show that TMO quality rankings strongly depend on the criteria used for the psychophysical evaluation. Not surprisingly, local TMOs are better than global TMOs on local criteria experiments and viceversa. We have found no significant correlation between local and global rankings, showing that observers are using several visual attributes to perform their tasks and some of these attributes are not considered by TMOs. We consider, these operators should take into account both local and global characteristics of the image and that there is ample room for improvement in the future development of TMO algorithms.  Furthermore, we suggest that an agreed standard criteria is needed for a proper and fair comparison among them.

Our rankings show there is no TMO that is clearly better than all the others across our experiments, but \textit{KimKautz} and \textit{Krawczyk} could be perhaps considered significantly better.

As a general conclusion, since currently there are no appropriate TMO for all situations, operators have to be selected depending on the observer's task.

\vskip21pt

\appendix

\section{Tone-Mapped Images}

\begin{figure*}[b!]
    \begin{subfigure}[b]{0.30\linewidth}
      \centering
      \centerline{\includegraphics[width=0.9\linewidth]{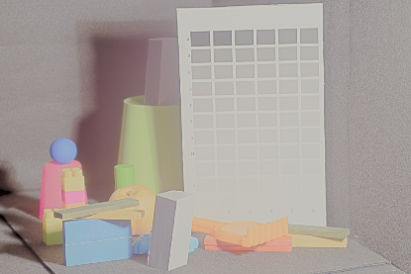}}
      \caption{\textit{Ashikhmin}}
    \end{subfigure}
    \hfill
    \begin{subfigure}[b]{0.30\linewidth}
      \centering
      \centerline{\includegraphics[width=0.9\linewidth]{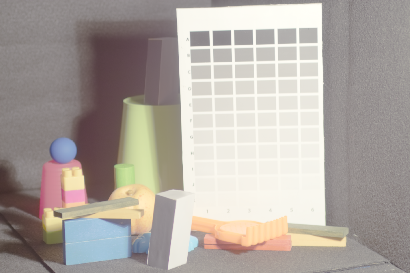}}
      \caption{\textit{Drago}}
    \end{subfigure}
    \hfill
    \begin{subfigure}[b]{0.30\linewidth}
      \centering
      \centerline{\includegraphics[width=0.9\linewidth]{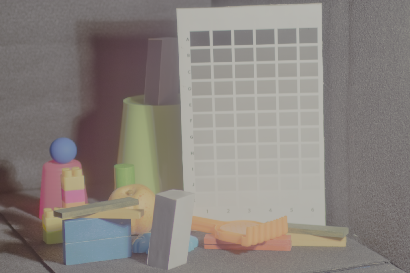}}
      \caption{\textit{Durand}}
    \end{subfigure}
    \hfill
    \begin{subfigure}[b]{.30\linewidth}
      \centering
      \centerline{\includegraphics[width=0.9\linewidth]{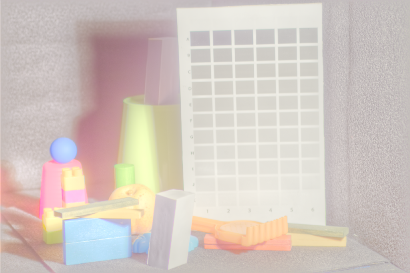}}
      \caption{\textit{Fattal}}
    \end{subfigure}
    \hfill
    \begin{subfigure}[b]{0.30\linewidth}
      \centering
      \centerline{\includegraphics[width=0.9\linewidth]{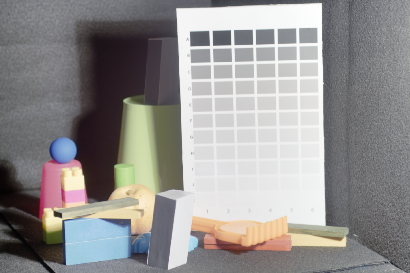}}
      \caption{\textit{Ferradans}}
    \end{subfigure}
    \hfill
    \begin{subfigure}[b]{0.30\linewidth}
      \centering
      \centerline{\includegraphics[width=0.9\linewidth]{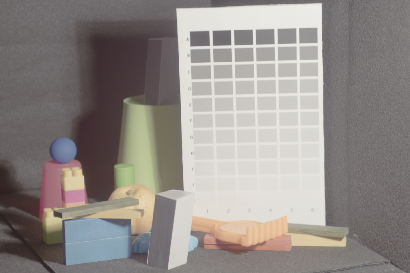}}
      \caption{\textit{Ferwerda}}
    \end{subfigure}
    \hfill
    \begin{subfigure}[b]{.30\linewidth}
      \centering
      \centerline{\includegraphics[width=0.9\linewidth]{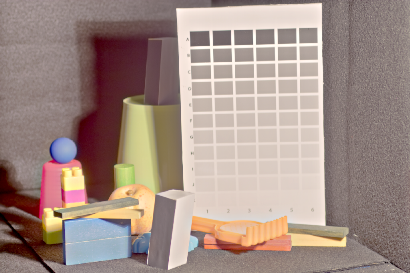}}
      \caption{\textit{iCAM06}}
    \end{subfigure}
    \hfill
    \begin{subfigure}[b]{0.30\linewidth}
      \centering
      \centerline{\includegraphics[width=0.9\linewidth]{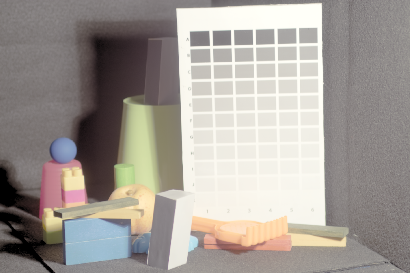}}
      \caption{\textit{KimKautz}}
    \end{subfigure}
    \hfill
    \begin{subfigure}[b]{0.30\linewidth}
      \centering
      \centerline{\includegraphics[width=0.9\linewidth]{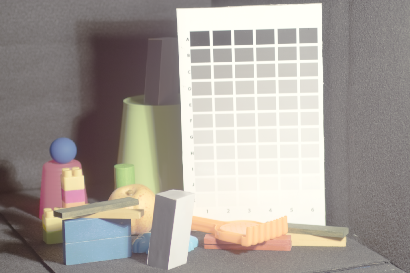}}
      \caption{\textit{Krawczyk}}
    \end{subfigure}
    \hfill
    \begin{subfigure}[b]{.30\linewidth}
      \centering
      \centerline{\includegraphics[width=0.9\linewidth]{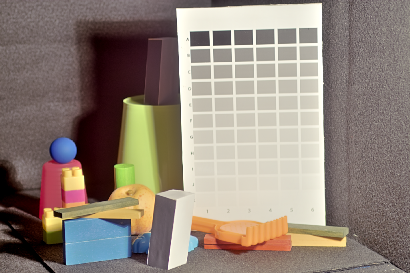}}
      \caption{\textit{Li}}
    \end{subfigure}
    \hfill
    \begin{subfigure}[b]{0.30\linewidth}
      \centering
      \centerline{\includegraphics[width=0.9\linewidth]{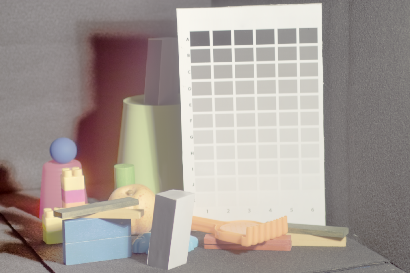}}
      \caption{\textit{Mertens}}
    \end{subfigure}
    \hfill
    \begin{subfigure}[b]{0.30\linewidth}
      \centering
      \centerline{\includegraphics[width=0.9\linewidth]{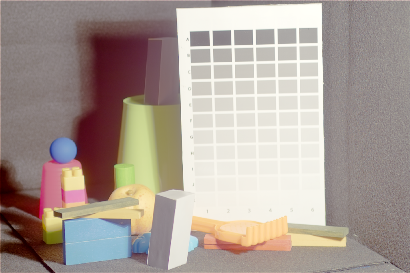}}
      \caption{\textit{Meylan}}
    \end{subfigure}
    \hfill
    \begin{subfigure}[b]{.30\linewidth}
      \centering
      \centerline{\includegraphics[width=0.9\linewidth]{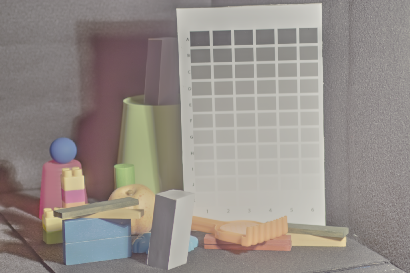}}
      \caption{\textit{Otazu}}
    \end{subfigure}
    \hfill
    \begin{subfigure}[b]{0.30\linewidth}
      \centering
      \centerline{\includegraphics[width=0.9\linewidth]{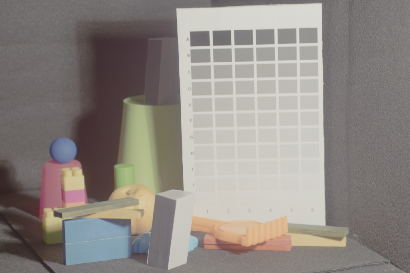}}
      \caption{\textit{Reinhard}}
    \end{subfigure}
    \hfill
    \begin{subfigure}[b]{0.30\linewidth}
      \centering
      \centerline{\includegraphics[width=0.9\linewidth]{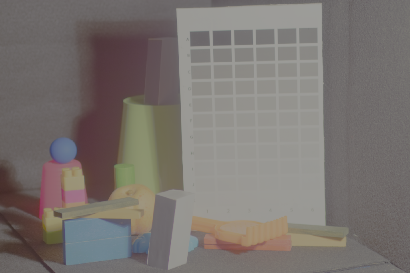}}
      \caption{\textit{Reinhard-Devlin}}
    \end{subfigure}
    \caption{Scene 1 Tone-mapped Images. For display purpose only, here all images are presented with a 2.2 gamma, except Reinhard-Devlin which is presented with a 1.6 gamma, as suggested by the authors.}
    \label{fig:scene1tonemapped}
\end{figure*}

\begin{figure*}[t!]
    \begin{subfigure}[b]{.30\linewidth}
      \centering
      \centerline{\includegraphics[width=0.9\linewidth]{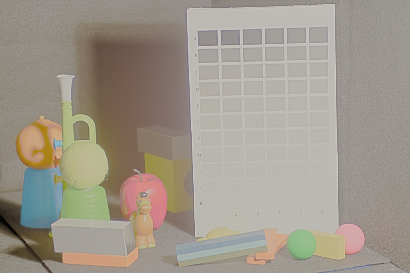}}
      \caption{\textit{Ashikhmin}}
    \end{subfigure}
    \hfill
    \begin{subfigure}[b]{0.30\linewidth}
      \centering
      \centerline{\includegraphics[width=0.9\linewidth]{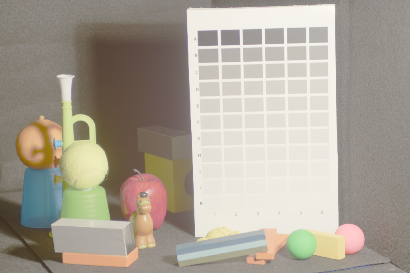}}
      \caption{\textit{Drago}}
    \end{subfigure}
    \hfill
    \begin{subfigure}[b]{0.30\linewidth}
      \centering
      \centerline{\includegraphics[width=0.9\linewidth]{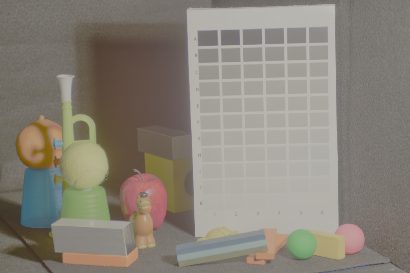}}
      \caption{\textit{Durand}}
    \end{subfigure}
    \hfill
    \begin{subfigure}[b]{.30\linewidth}
      \centering
      \centerline{\includegraphics[width=0.9\linewidth]{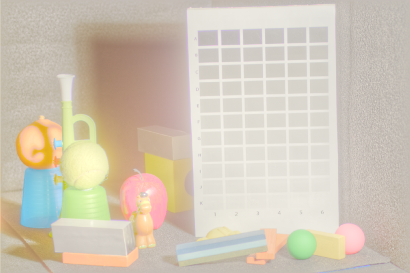}}
      \caption{\textit{Fattal}}
    \end{subfigure}
    \hfill
    \begin{subfigure}[b]{0.30\linewidth}
      \centering
      \centerline{\includegraphics[width=0.9\linewidth]{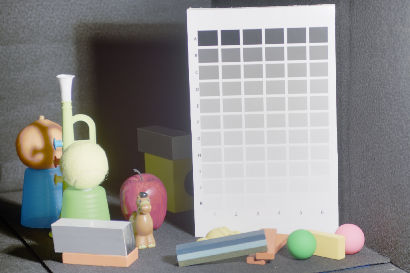}}
      \caption{\textit{Ferradans}}
    \end{subfigure}
    \hfill
    \begin{subfigure}[b]{0.30\linewidth}
      \centering
      \centerline{\includegraphics[width=0.9\linewidth]{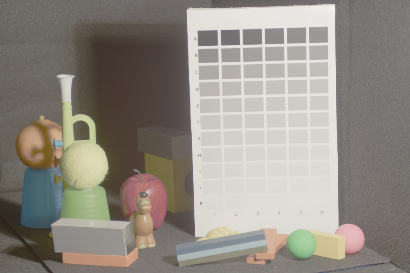}}
      \caption{\textit{Ferwerda}}
    \end{subfigure}
    \hfill
    \begin{subfigure}[b]{.30\linewidth}
      \centering
      \centerline{\includegraphics[width=0.9\linewidth]{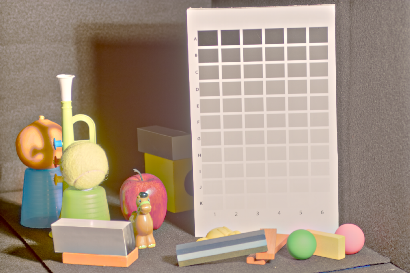}}
      \caption{\textit{iCAM06}}
    \end{subfigure}
    \hfill
    \begin{subfigure}[b]{0.30\linewidth}
      \centering
      \centerline{\includegraphics[width=0.9\linewidth]{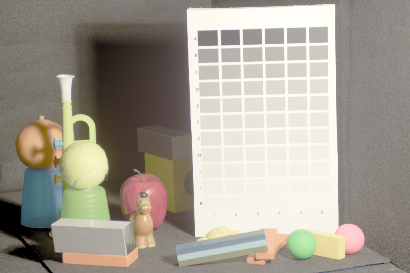}}
      \caption{\textit{KimKautz}}
    \end{subfigure}
    \hfill
    \begin{subfigure}[b]{0.30\linewidth}
      \centering
      \centerline{\includegraphics[width=0.9\linewidth]{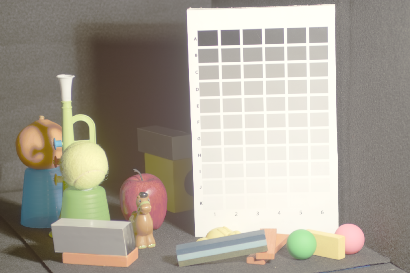}}
      \caption{\textit{Krawczyk}}
    \end{subfigure}
    \hfill
    \begin{subfigure}[b]{.30\linewidth}
      \centering
      \centerline{\includegraphics[width=0.9\linewidth]{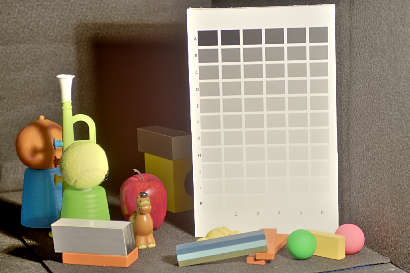}}
      \caption{\textit{Li}}
    \end{subfigure}
    \hfill
    \begin{subfigure}[b]{0.30\linewidth}
      \centering
      \centerline{\includegraphics[width=0.9\linewidth]{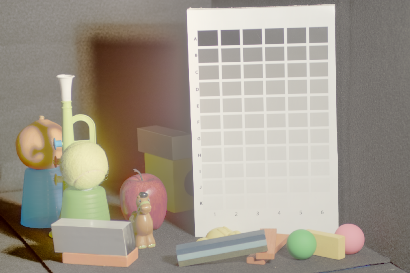}}
      \caption{\textit{Mertens}}
    \end{subfigure}
    \hfill
    \begin{subfigure}[b]{0.30\linewidth}
      \centering
      \centerline{\includegraphics[width=0.9\linewidth]{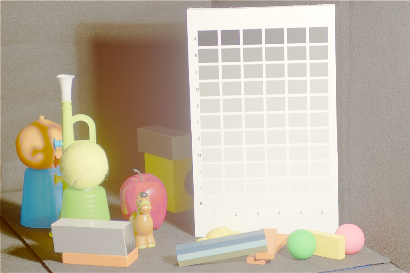}}
      \caption{\textit{Meylan}}
    \end{subfigure}
    \hfill
    \begin{subfigure}[b]{.30\linewidth}
      \centering
      \centerline{\includegraphics[width=0.9\linewidth]{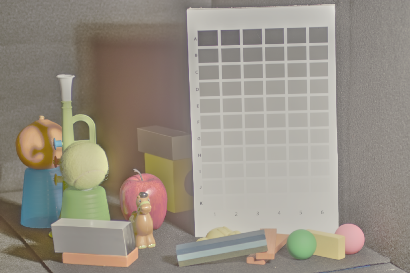}}
      \caption{\textit{Otazu}}
    \end{subfigure}
    \hfill
    \begin{subfigure}[b]{0.30\linewidth}
      \centering
      \centerline{\includegraphics[width=0.9\linewidth]{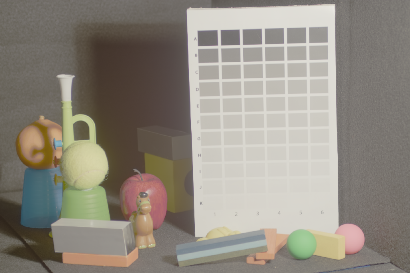}}
      \caption{\textit{Reinhard}}
    \end{subfigure}
    \hfill
    \begin{subfigure}[b]{0.30\linewidth}
      \centering
      \centerline{\includegraphics[width=0.9\linewidth]{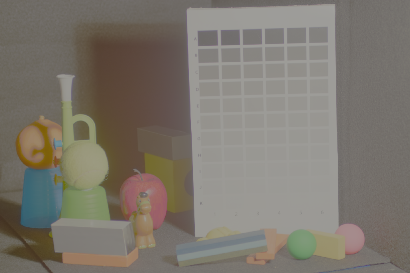}}
      \caption{\textit{Reinhard-Devlin}}
    \end{subfigure}
    \caption{This figure is the same as previous one, but showing scene 2 tone-mapped images.}
    \label{fig:scene2tonemapped}
\end{figure*}

\begin{figure*}[t!]
    \begin{subfigure}[b]{.30\linewidth}
      \centering
      \centerline{\includegraphics[width=0.9\linewidth]{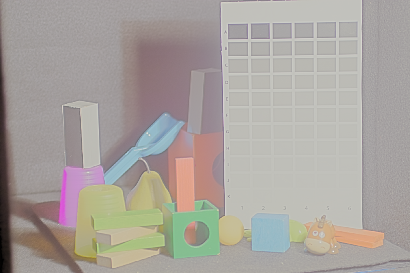}}
      \caption{\textit{Ashikhmin}}
    \end{subfigure}
    \hfill
    \begin{subfigure}[b]{0.30\linewidth}
      \centering
      \centerline{\includegraphics[width=0.9\linewidth]{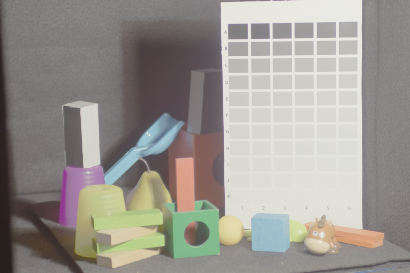}}
      \caption{\textit{Drago}}
    \end{subfigure}
    \hfill
    \begin{subfigure}[b]{0.30\linewidth}
      \centering
      \centerline{\includegraphics[width=0.9\linewidth]{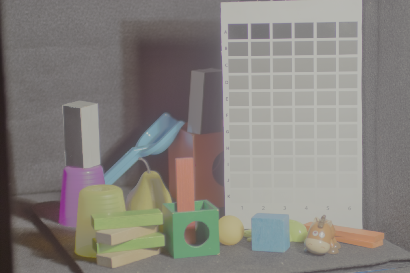}}
      \caption{\textit{Durand}}
    \end{subfigure}
    \hfill
    \begin{subfigure}[b]{.30\linewidth}
      \centering
      \centerline{\includegraphics[width=0.9\linewidth]{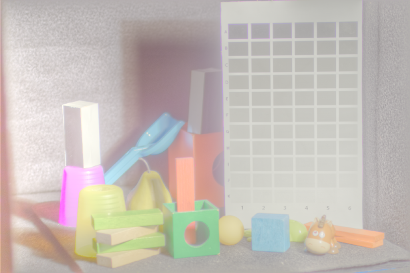}}
      \caption{\textit{Fattal}}
    \end{subfigure}
    \hfill
    \begin{subfigure}[b]{0.30\linewidth}
      \centering
      \centerline{\includegraphics[width=0.9\linewidth]{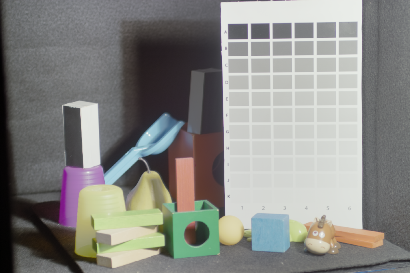}}
      \caption{\textit{Ferradans}}
    \end{subfigure}
    \hfill
    \begin{subfigure}[b]{0.30\linewidth}
      \centering
      \centerline{\includegraphics[width=0.9\linewidth]{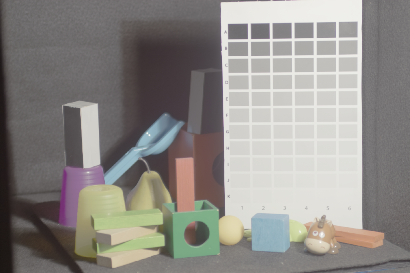}}
      \caption{\textit{Ferwerda}}
    \end{subfigure}
    \hfill
    \begin{subfigure}[b]{.30\linewidth}
      \centering
      \centerline{\includegraphics[width=0.9\linewidth]{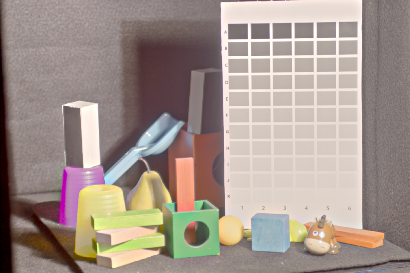}}
      \caption{\textit{iCAM06}}
    \end{subfigure}
    \hfill
    \begin{subfigure}[b]{0.30\linewidth}
      \centering
      \centerline{\includegraphics[width=0.9\linewidth]{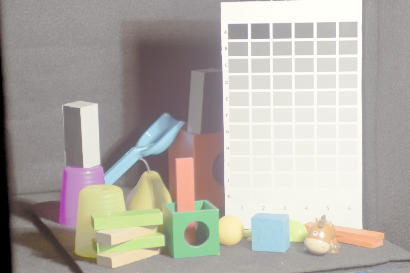}}
      \caption{\textit{KimKautz}}
    \end{subfigure}
    \hfill
    \begin{subfigure}[b]{0.30\linewidth}
      \centering
      \centerline{\includegraphics[width=0.9\linewidth]{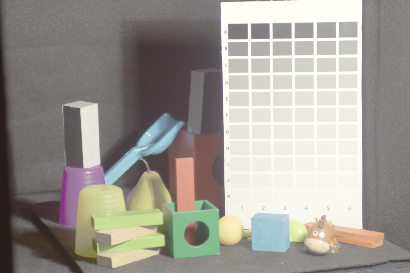}}
      \caption{\textit{Krawczyk}}
    \end{subfigure}
    \hfill
    \begin{subfigure}[b]{.30\linewidth}
      \centering
      \centerline{\includegraphics[width=0.9\linewidth]{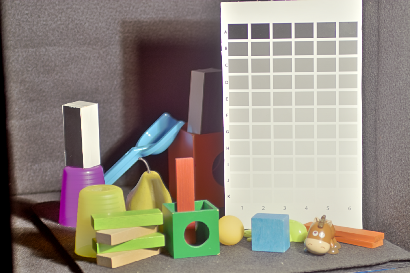}}
      \caption{\textit{Li}}
    \end{subfigure}
    \hfill
    \begin{subfigure}[b]{0.30\linewidth}
      \centering
      \centerline{\includegraphics[width=0.9\linewidth]{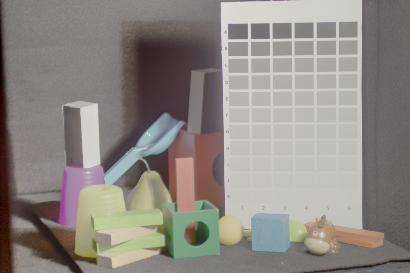}}
      \caption{\textit{Mertens}}
    \end{subfigure}
    \hfill
    \begin{subfigure}[b]{0.30\linewidth}
      \centering
      \centerline{\includegraphics[width=0.9\linewidth]{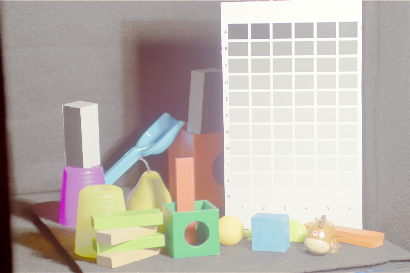}}
      \caption{\textit{Meylan}}
    \end{subfigure}
    \hfill
    \begin{subfigure}[b]{.30\linewidth}
      \centering
      \centerline{\includegraphics[width=0.9\linewidth]{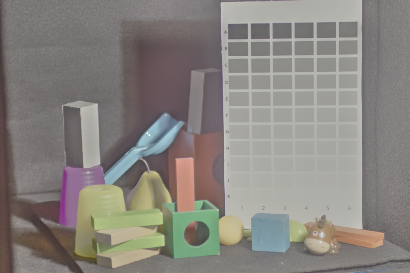}}
      \caption{\textit{Otazu}}
    \end{subfigure}
    \hfill
    \begin{subfigure}[b]{0.30\linewidth}
      \centering
      \centerline{\includegraphics[width=0.9\linewidth]{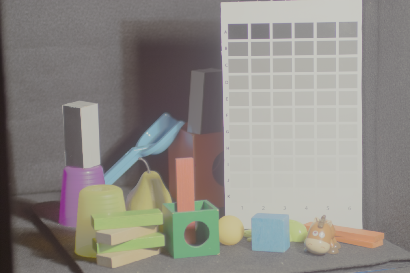}}
      \caption{\textit{Reinhard}}
    \end{subfigure}
    \hfill
    \begin{subfigure}[b]{0.30\linewidth}
      \centering
      \centerline{\includegraphics[width=0.9\linewidth]{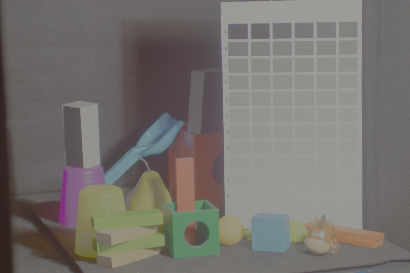}}
      \caption{\textit{Reinhard-Devlin}}
    \end{subfigure}
    \caption{This figure is the same as previous one, but showing scene 3 tone-mapped images.}
    \label{fig:scene3tonemapped}
\end{figure*}

In Figures~\ref{fig:scene1tonemapped} to~\ref{fig:scene3tonemapped} all the tone-mapped images of each scene used in the experiments are shown.

\section{Statistical Results}
\label{Appendix:statisticalResults}
\begin{figure*}
    \centering
    \includegraphics[width=\textwidth]{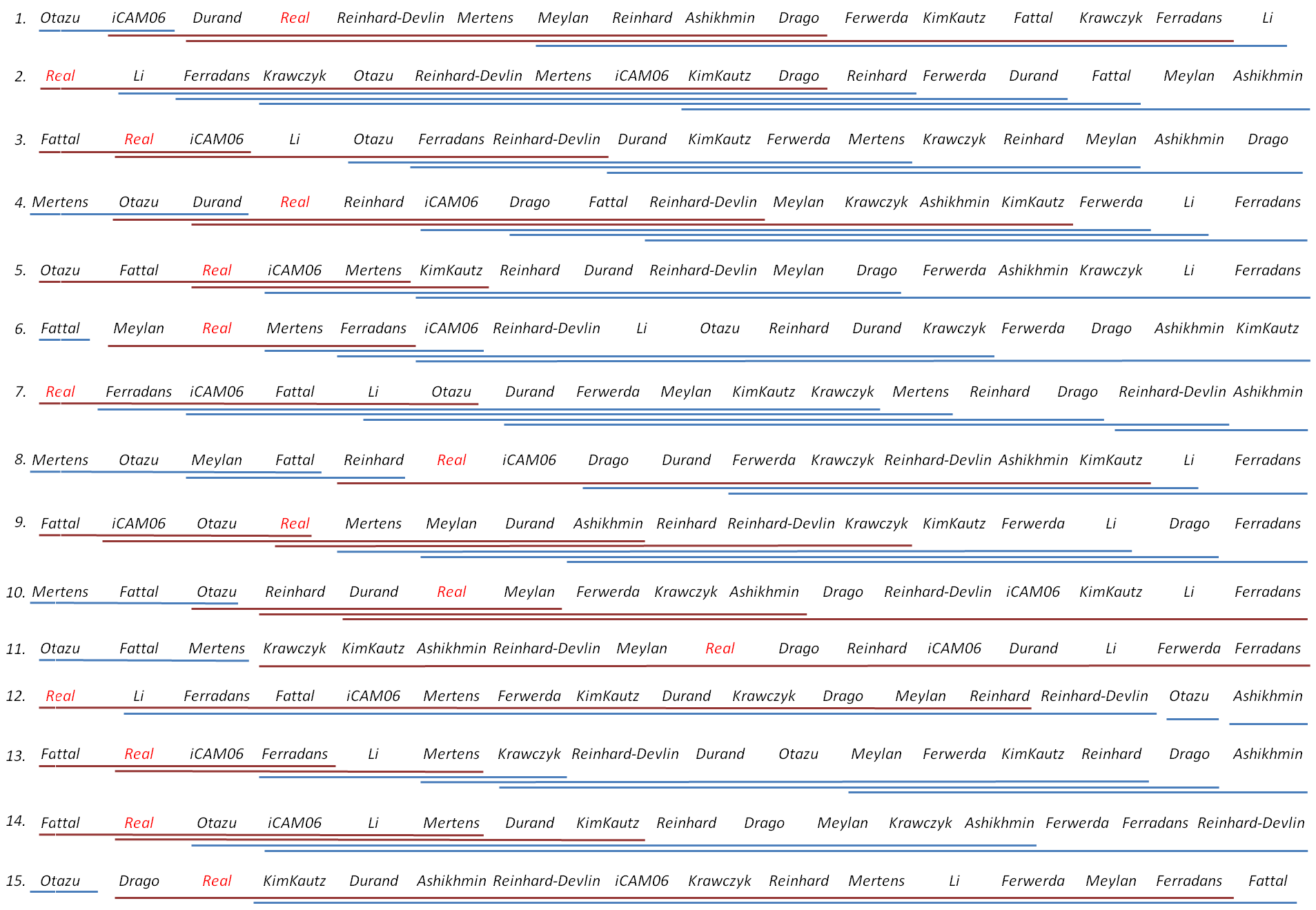}
    \caption{Fisher's Least Significant Difference Post-hoc Test carried out for all grey-level surfaces. Each line corresponds to the set of TMOs that are not significantly different at $p<0.05$. The real scene is indicated in red.}
    \label{fig:totalFLSD}
\end{figure*}

In this section we show all the statistical results used to construct the different rankings of Section~\ref{Experim1:results}. Table~\ref{table:FpANOVA} shows F and p values from the ANOVA analysis. Figure~\ref{fig:totalFLSD} shows results of Fisher's Least Significant Difference Post-hoc test.

\begin{table}[ht]
        \centering
        \begin{tabular}{ccc}
            \hline
            \multicolumn{3}{c}{ANOVA analysis} \\
            \hline
            Surface & F  & p \\
            \hline
            1   & 2.74 & $0.001$ \\
            2   & 3.11 & $< 0.001$ \\
            3   & 5.74 & $< 0.001$ \\
            4   & 3.31 & $< 0.001$ \\
            5   & 3.70 & $< 0.001$ \\
            6   & 6.08 & $< 0.001$ \\
            7   & 3.65 & $< 0.001$ \\
            8   & 6.77 & $< 0.001$ \\
            9   & 3.25 & $< 0.001$ \\
            10  & 3.67 & $< 0.001$ \\
            11  & 4.33 & $< 0.001$ \\
            12  & 7.12 & $< 0.001$ \\
            13  & 5.59 & $< 0.001$ \\
            14  & 2.28 & $< 0.001$ \\
            15  & 4.49 & $< 0.001$ \\
            \hline
        \end{tabular}
        \caption{F and p values obtained from ANOVA analysis for each evaluated surface.}
        \label{table:FpANOVA}
\end{table}

\begin{section}{Acknowledgements}
C. Alejandro Parraga and Xavier Otazu have been partially supported by the Spanish Ministry of Science and Innovation through research projects TIN2013-41751 and TIN2013-49982-EXP.

We would like to thank Carlo Gatta for his useful comments on the psychophysical experiments design and Javier Retana for his useful comments on statistical analysis procedures.

Thanks to all subjects who have participated in the psychophysical experiments, and all authors who publicly share their code.

\end{section}

\bibliographystyle{apalike}
\bibliography{refs}

\end{document}